\documentclass[aps,prd,preprint,superscriptaddress,showpacs,nofootinbib]{revtex4-1}
\usepackage{latexsym,axodraw}
\usepackage{amssymb, amsmath, slashed}
\usepackage{epsfig, graphics,pstricks, color}
\numberwithin{equation}{section}

\newcommand{\be}{\begin{equation} } \newcommand{\ee}{\end{equation} } 
\newcommand{\bear}{\begin{eqnarray} } \newcommand{\eear}{\end{eqnarray} }
\newcommand{\gev}{\text{GeV}}
\newcommand{\tev}{\text{TeV}}
\newcommand{\jets}{\text{jets}}

\newcommand{\fb}{\text{fb}}

\begin{document}

\baselineskip=21pt \pagestyle{plain} \setcounter{page}{1}

\vspace*{-1.7cm}

\begin{flushright}{\small Fermilab-PUB-14-264-T}\end{flushright}

\vspace*{0.2cm}
 
\begin{center}

{\Large \bf Interpretations of anomalous LHC events \\ [3mm] with electrons and jets}\\ [9mm]

{\normalsize \bf Bogdan A. Dobrescu$^\star$ and Adam Martin$^\diamond$ \\ [4mm]
$\star$ {\it Theoretical Physics Department, Fermilab, Batavia, IL 60510, USA } \\ [1mm]
$\diamond${ \it Department of Physics, 
University of Notre Dame, Notre Dame, IN 46556, USA}

}

\vspace*{0.5cm}

August 5, 2014; revised September 7, 2014

\vspace*{0.7cm}

{\bf \small Abstract}

\vspace*{0.4cm}

\parbox{14.8cm}{
The CMS Collaboration has recently reported some excess events in final states with electrons and jets, in searches for leptoquarks and 
$W'$ bosons. Although these excesses may be due to some yet-to-be-understood  background mismodeling, it is 
useful to seek realistic interpretations involving new particles that could generate such events.
We show that resonant pair production of vector-like leptons that decay to an electron and two jets leads to kinematic distributions 
consistent with the CMS data. 
}

\end{center}

\nopagebreak
\section{Introduction}

The success of the Standard Model (SM) in describing the large number of 
final states analyzed so far at the LHC is remarkable.  At the same time, there are many 
hypothetical particles whose signatures have not been searched for yet, especially when 
two or more of these particles are produced in the same process. 
An example is provided by the pair production of some new heavy particle
whose decays are controlled by an approximate symmetry. The dominant decays of such a particle may involve 
peculiar flavor combinations of many SM particles, while the constraints from low-energy flavor 
processes are avoided if the breaking of the symmetry that controls the decay parameters is sufficiently soft.
Nevertheless, the ensuing signals could show up in some of the existent searches, even though the event selection 
is not chosen for those signals.

Recently, the CMS Collaboration has reported excess events involving electrons and jets, while similar events involving 
muons and jets are well described by the SM. In the search for a $W'$ boson decaying to the $eejj$ final state \cite{Khachatryan:2014dka}, 
the leading two electrons and two jets form an invariant mass
peak near 2.1 TeV; even though the statistical significance is only $2.8 \sigma$, this signal is sufficiently clean to warrant further scrutiny.
Based on some kinematic distributions, the CMS Collaboration concluded that the signal is not consistent with the $W'$ hypothesis.

In the CMS search for first-generation leptoquarks \cite{CMS:2014qpa} the leading two electrons and two jets pair up so that the $ej$ invariant mass exhibits
a broad excess with a maximum at about 500 GeV.
The statistical significance of  this excess is $2.4\sigma$ when a cut on the invariant mass of the $ej$ system with smaller $m_{ej}$
 is imposed ($m_{ej}^{\rm min} > 360$ GeV). When that cut is removed, the deviation from the SM increases substantially, despite larger systematic uncertainties.
The leptoquark hypothesis would imply a very narrow $ej$ invariant mass  peak, which is 
again not consistent with the observed excess. 

The leptoquark search \cite{CMS:2014qpa} has also led to 
an excess of events with one electron, two jets and missing transverse energy ($\slash \!\!\!\! E_T$), roughly consistent with the 
hypothesis that one leptoquark decays into $ej$ and the other one into $\nu j $. This $e\nu jj$ signal has a statistical significance 
of $2.6\sigma$, and entails a broad $ej$ peak at around 700 GeV.

Given that the event selections are sufficiently complicated (especially for the leptoquark searches),
it may be that there are correlations between the events that contribute to these three excesses, in which case the combined
statistical significance my not be as large as it seems. 
It is also possible that the background modeling is not sufficiently accurate, but this would be surprising for 
the very high $p_T$ events involved in the $W'$ and leptoquark searches.

Here we investigate the possibility that the $ee jj$ and $e\nu jj$ excesses are due to some new particles. 
The resonant production of leptoquarks, investigated in Ref.~\cite{Bai:2014xba}, leads to an $eejj$ invariant mass peak but does not appear to 
be consistent with the broad $ej$ peaks. By contrast, our starting point is resonant signatures involving 
four (or more) jets plus an $e^+e^-$ pair. Such  final states lead to combinatoric ambiguities due to multiple choices for the leading two jets selected by 
the CMS leptoquark search. As a result the $ej$ invariant mass distributions become very wide. Furthermore, we show that although the $ee+2j$ 
invariant mass peak has a flatter distribution within these $ee+4j$ final states compared to a resonant $eejj$ signal, the remaining edge-like feature provides a good fit to the excess in the CMS $W'$ search. 

In Section \ref{sec:selections} we describe the details of the CMS searches with electrons and jets, and select the type of particles that may be responsible for the
observed excesses.
In Section \ref{sec:models} we propose a few models involving a $Z'$ boson that is produced in the $s$ channel and decays into a pair of heavy vector-like leptons.
The decays of the vector-like leptons lead to signatures that may explain the CMS events, as shown in Section \ref{sec:sim}. 
We discuss additional tests of these models, and then in Section \ref{sec:Conclusions} provide a brief outlook.

\section{Event selections and signal features}
\label{sec:selections}

We start by reviewing the analysis cuts imposed 
in each of the three CMS searches with electrons and jets, and then by drawing some conclusions about the type of new particles that could cause the disagreement between data and SM expectation.

In the first-generation leptoquark searches \cite{CMS:2014qpa}, both the single- and di-electron searches require at least two jets, $p_T > 45\,\gev, |\eta| < 2.4$, with the leading jet satisfying $p_T > 125\,\gev$.
The analysis uses the anti-$k_T$ jet algorithm with size $R =0.5$. A minimum jet-electron separation of $0.3$ is also imposed. Electron candidates must have $p_T > 45\,\gev$, and $|\eta| < 2.5$; exactly two candidates are required for the di-electron channel, while in the single electron channel there is an additional requirement of large missing energy that is azimuthally well separated from the electron and leading jet: $\slashed E_T > 55\,\gev$, $\Delta \phi(\slashed E_T, e) > 0.8, \Delta \phi(\slashed E_T,j) > 0.5$. In both channels, any events containing a muon $p_T > 10\,\gev$ are vetoed. 

Further cuts, chosen to optimize the analyses for a given leptoquark mass, are subsequently applied in each channel. For the di-electron channel, the cuts used for optimization are the scalar sum of the $p_T$ of the two leptons and the leading two jets $S_T$, the di-electron mass $m_{ee}$ and the minimum electron-jet invariant mass $m^{\rm min}_{ej}$. The $m^{\rm min}_{ej}$ is calculated by first pairing each electron in the event with one of the hardest two jets; there are two ways to pair the objects and the choice that is most consistent with production of a pair of equal-mass $ej$ resonances is made. Once the pairing has been decided, the minimum mass of the two $ej$ systems becomes $m^{\rm min}_{ej}$. The optimization
cuts for a 650 GeV leptoquark signal are $S_T > 850\,\gev,\, m_{ee} > 155\,\gev$ and $m^{\rm min}_{ej} > 360\,\gev$. For these cut values, $36$ events are observed compared with $20.49\,\pm2.4\,\pm2.45\text{(syst.)}$ events expected from SM backgrounds ($Z + \jets,\ t\bar t$, and multijet QCD). 
The excess has been examined for jet-flavor content and does not appear to contain $b$-jets.
When the $m^{\rm min}_{ej}$ cut is removed, Figure 11 of Ref.~\cite{CMS:2014qpa} shows that there are additional 49 events observed for a background 
in the range of 23--33 events.

The fact that the $m^{\rm min}_{ej}$ distribution is so broad suggests that there is no on-shell particle decaying to $ej$. In Section \ref{sec:models} 
we construct a model
where a very heavy scalar leptoquark, integrated out, induces 3-body decays of a parent fermion, leading to a broad  $m^{\rm min}_{ej}$ distribution.
To have the peak of that distribution at $\sim 500$ GeV it is necessary for the new fermion (which decays to $ejj$) to be somewhat heavier than that. 
We  also construct two models where there is no particle coupled to an electron and a jet, so that any pairing of one electron and one jet in the final state
is non-resonant; a broad  $m^{\rm min}_{ej}$  peak is instead induced by cuts and kinematic limits.

In the $e\nu jj$  
analysis of the leptoquark search \cite{CMS:2014qpa}, the optimization cuts are on the missing energy $\slashed E_T$, $S_T$ (defined now 
as the scalar sum of the missing energy  with the $p_T$ of the electron and of the two leading jets), the mass of the electron-jet system $m_{ej}$,
 and the transverse mass of the $\slashed E_T$-electron system $m_{T,e\nu}$. As with $m^{\rm min}_{ej}$ above, there are two possible choices for $m_{ej}$. Here, the choice that minimizes $|m_{T, ej} - m_{T,\nu j}|$ is used. For the same leptoquark signal assumption as in the di-electron channel, the optimization cut
 values are $S_T > 1040\, \gev$, $\slashed E_T > 145\,\gev$, $m_{ej} > 555\,\gev$ and $m_{T,e\nu} > 270\,\gev$, under which 18 events are observed with $7.54\,\pm 1.20\,\pm 1.07\text{(syst.)}$ expected. 
 
It is intriguing that the peak of the $m_{ej}$ distribution in the $e\nu jj$ search is at a value comparable to that of the $m^{\rm min}_{ej}$ distribution in the 
``leptoquark" $eejj$ search. This suggests that one of the electrons observed in the $eejj$ search originates from a vertex involving 
the lepton doublet $(\nu_L^e, e_L)$. A further supporting fact is that the cross sections for the two excesses are comparable, of order 1 fb.

 The excess di-electron events observed in the $W'\to eejj$ search \cite{Khachatryan:2014dka}  
 have been subjected to a different, though not orthogonal set of analysis cuts. There, the electron cuts are staggered: $p_T > 60\,\gev, |\eta| < 2.5$ ($p_T  > 40\,\gev, |\eta| < 2.5$) for the leading (subleading) electron. At least two jets are required, as before, though at a slightly lower $p_T$ threshold of $40\,\gev$. For all events that satisfy $m_{ee} > 200\,\gev$, the two electrons and two leading jets are combined, and the total invariant mass is calculated. While $m_{eejj}$ data and the SM prediction agree at lower values, there is an excess of events quoted as $2.8\,\sigma$ deviation in the $1.8\,\tev  < m_{eejj} < 2.2\, \tev$ bin (roughly $14$ events observed with $4$ expected). 
The similar ATLAS search for $W' \to \ell N \to \ell\ell jj$ used only 2.1 fb$^{-1}$ of 7 TeV data \cite{ATLAS:2012ak}, and is not sensitive to masses of 2 TeV.

The concentration of events in one high-mass bin strongly suggests a resonant production of the entire final state. While a new spin-1 particle 
(a $W'$ boson as in \cite{Khachatryan:2014dka}, a coloron as in \cite{Bai:2014xba}, or a $Z'$ boson as in our models presented in Section \ref{sec:models}) 
with sizable couplings to the quark-antiquark initial states can easily generate the required 2 TeV peak in $m_{eejj}$, the fit to additional 
kinematic distributions is difficult to achieve. 
The $W'\to e N \to eejj$ hypothesis, although well motivated \cite{Heikinheimo:2014tba}, is not 
consistent with some (unspecified) kinematic distributions, as mentioned by the CMS Collaboration \cite{Khachatryan:2014dka}.
It is then useful to try all possible decay patterns leading to $eejj$ final states \cite{Abdullah:2014oaa}. In our case, given that we 
try to find a common origin with the signal seen in the leptoquark search, we attempt to find models where the signal responsible 
for the $m_{eejj}$ peak gives an $m^{\rm min}_{ej}$ distribution as in \cite{CMS:2014qpa}. To that end we have analyzed a model 
where $Z'\to \bar{q} \chi$, where $\chi$ is a vector-like quark decaying to $e^+e^-q$; the $m^{\rm min}_{ej}$ distribution turns out 
to be too flat in that case, with a peak at a $m^{\rm min}_{ej}$ value that is too large, almost independently of the $\chi$ mass.

A more promising hypothesis, consistent with the features of the ``leptoquark signal", is that the $s$-channel resonance decays 
to $e^+e^-$ and more than two jets. That can be consistent with the ``$W'$ signal" because the $1.8\,\tev  < m_{eejj} < 2.2\, \tev$ bin where the events are concentrated  is wide enough. In Section \ref{sec:sim} we will show that $ee+4j$ final states, arising from the $Z'$ models presented next,
indeed satisfy this requirement. Given that some of the energy released in the $Z'$ decay is taken away by the third and fourth jets,
the $Z'$ mass has to be slightly larger than 2.2 TeV.


\section{Resonant production of vector-like leptons}\label{sec:models}

Let us consider the SM plus a $U(1)_B$ gauge group, spontaneously broken by the VEV of a scalar $\phi_B$. 
The new heavy gauge boson, $Z^\prime$, has a mass $M_{Z'}$, which should be in the $2.2 - 2.5$ TeV range 
in order to produce a resonant signal consistent with the CMS $eejj$ excess. 
We impose that all SM quarks have same $U(1)_B$ charge (chosen to be 1/3), and all SM leptons are $U(1)_B$ neutral, so that 
the constraints from FCNCs and di-lepton resonance searches are avoided.
Besides the SM fermions we include a vector-like lepton that carries $U(1)_B$ charge $z_{vl}$ and is an $SU(2)_W$ singlet. (More generally, the $U(1)_B$
charges of its left-and right-handed components could differ, but this possibility would not modify our conclusions;  in either case,
the anomaly cancellation conditions 
for $U(1)_B$ require additional vector-like fermions \cite{Dobrescu:2013cmh, Dobrescu:2014fca,Carena:2004xs}.)

We consider the cases where the vector-like fermion is electrically neutral and is labelled by $N$, or carries electric charge $-1$ and is labelled by $E$.
We assume that its mass ($m_N$ or $m_E$) is less than half the $Z'$ mass $M_{Z'}$, and that its Yukawa couplings to the SM leptons and the Higgs doublet 
are negligible.
The normalization for $g_z$ used in what follows is that where the $Z'$ couplings to SM quarks are given by $(g_z/6) Z'_\mu \bar q \gamma^\mu q$.

The branching fraction for $Z' \to N\bar{N}$ (or for $Z' \to E^+ E^-$, with $m_E$ replacing $m_N$) is
\be
B(Z' \to N\bar{N}) = \frac{z_{vl}^2 \beta_N}{2 (1+\alpha_s/\pi) + z_{vl}^2 \beta_N}  ~~,
\ee
where $\beta_N = (1-4m_N^2/M_{Z'}^2)^{1/2}$, $\alpha_s$ is the strong coupling constant, and we have ignored electroweak corrections. 
For concreteness, we will use $z_{vl}=1$. 

For  $M_{Z'} = 2.3$ TeV and  $m_N = 700$ GeV we find
$B(Z' \to N\bar{N}) \approx 26\%$, and the leading-order cross sections (computed with MadGraph 5 \cite{Alwall:2014hca})
for $N\overline N$ production at the LHC with $\sqrt{s} = (8,13, 14)$ TeV are $g_z^2 \times (7.3, 47, 61)$ fb,
respectively. 
For  $M_{Z'} = 2.4$ TeV and  $m_E = 800$ GeV we find
$B(Z' \to E^+E^-) \approx 25\%$, and the leading-order cross sections
for $E^+E^-$ LHC production with $\sqrt{s} = (8,13, 14)$ TeV are $g_z^2 \times (3.6, 9.3, 12)$ fb,
respectively. 

QCD effects at next-to-leading order (NLO)  typically increase the $Z'$ production by about 30\%. Given that we use only leading order production 
cross section, the NLO effects may be taken into account through a reduction in $g_z$ by about 15\%.

An upper limit on $g_z$ follows from di-jet resonance searches; for $M_{Z'} = 2.3$ TeV, the limit derived as in \cite{Dobrescu:2013cmh}
is given by 
\be
g_z \left( 1 - B(Z' \to N\bar{N}) \right)^{1/2} < 1.6 ~~,
\ee
based on the ATLAS search with 20.3 fb$^{-1}$ \cite{Aad:2014aqa}. With $B(Z' \to N\bar{N}) = 26\%$, the upper limit on the 
$Z'$ gauge coupling is $g_z < 1.8$. For $M_{Z'} = 2.4$ TeV the limit is weaker.

There are several possibilities for the decays of $N$ or $E$. We concentrate on three models, described in what follows.

\subsection{$Z'N$ Model}

In the first one, referred to as the $Z'N$ Model,  $N$ has 3-body decays into $e^- u\bar{d}$ or $\nu d \bar{d}$  via the dimension-7 operator
\be
\frac{y_{Nq} y_{dl}}{M_{\rm LQ}^3} 
\phi_B (\bar{\cal Q}_L^1 N_R)i\sigma_2 (\bar{\cal L}_L^1 d_R)  ~~,
\label{dim-6}
\ee
where ${\cal Q}_L^1 = (u_L,d_L)$ and ${\cal L}_L^1 = (\nu^e_L, e_L)$ are the quark and lepton doublets of the first generation,
$M_{\rm LQ} \gg M_N$ is related to the mass of a particle whose exchange generates the dimension-7 operator, and 
$y_{Nq}$, $y_{dl}$ are its couplings to fermions. We assigned $U(1)_B$ charge $-1$ to the scalar $\phi_B$ responsible for spontaneous $U(1)_B$ breaking. 

A renormalizable UV completion is provided by 
two scalar leptoquarks, $\tilde{q}_0$ and $\tilde{q}_1$, which carry $U(1)_B$ charge 0 and $+1$ respectively, and have a trilinear term $\tilde{q}_1 \tilde{q}_0^\dagger  \phi_B$.
The Yukawa couplings $y_{Nq} \, \tilde q_0 \bar{\cal L}_L^1 d_R$ and $y_{dl} \, \tilde q_1^\dagger  \bar{\cal Q}_L^1 N_R$ then induce the operator (\ref{dim-6}) at tree level, once the heavy 
$\tilde{q}_0$ and $\tilde{q}_1$ are integrated out.

The flavor structure of the operator can be approximately enforced 
by a discrete symmetry that distinguishes the first generation of quark and leptons. Such a symmetry is broken by the mixing of the 
quark doublets required to generate the CKM matrix, but this effect is not problematic from the point of view of FCNCs because 
$y_{Nq}$ and $y_{dl}$ can be very small. The only requirement for producing a large signal with electrons and jets is that 
 the 3-body decays of $N$ into an electron and a quark-antiquark pair of the first or second generation has a large enough  branching fraction.

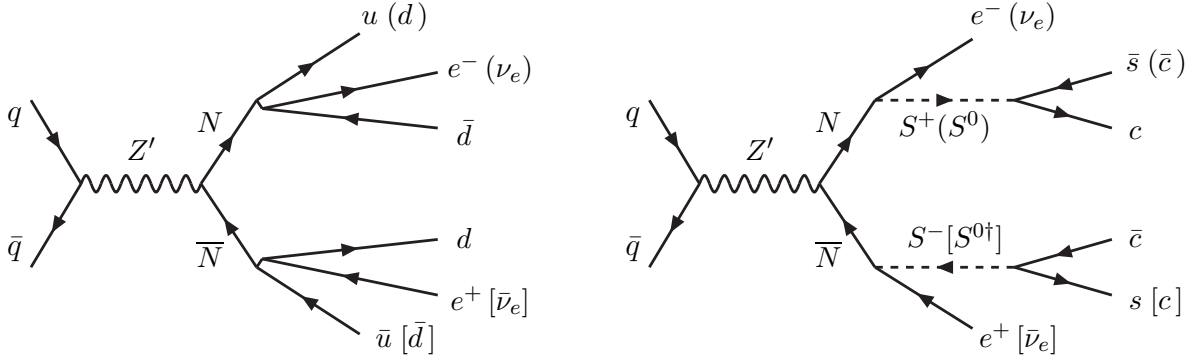
\begin{figure}[t]
\begin{center} 
{
\unitlength=2. pt
\SetScale{1.05}
\SetWidth{1.}      
\normalsize    
{} \allowbreak
\begin{picture}(100,100)(7,-37)
\ArrowLine(4,80)(22,50)
\ArrowLine(22,50)(4,20)
\Photon(22,50)(65,50){3}{6}
\ArrowLine(65,50)(85,80)
\ArrowLine(85,20)(65,50)
\ArrowLine(85,80)(122,104)\ArrowLine(87,77)(150,90)\ArrowLine(150,70)(87,77)\DashLine(85,80)(87,77){3}
\ArrowLine(122,-4)(85,20)\ArrowLine(87,23)(150,30)\ArrowLine(150,10)(87,23)\DashLine(85,20)(87,23){3}
\Text(-1,39)[c]{$q$}\Text(-1,14)[c]{$\bar q$}\Text(23,33)[c]{\small $ Z'$}
\Text(36,38)[c]{\small $N$}\Text(36,13)[c]{\small $\overline N$}
\Text(71,58)[c]{\small $u \; (d\,)$}\Text(89,48)[c]{\small $e^- \, (\nu_e)$}\Text(84,36)[c]{\small $\bar d$}
\Text(73,-3)[c]{\small $\bar u \; [\bar d\,]$}\Text(89,4)[c]{\small $e^+ \, [\bar\nu_e]$}\Text(84,16)[c]{\small $ d$}
\end{picture}
\begin{picture}(100,100)(-8,-37)
\ArrowLine(4,80)(22,50)
\ArrowLine(22,50)(4,20)
\Photon(22,50)(65,50){3}{6}
\ArrowLine(65,50)(85,80)
\ArrowLine(85,20)(65,50)
\ArrowLine(85,80)(120,102)\DashArrowLine(85,80)(135,80){3}\ArrowLine(170,90)(135,80)\ArrowLine(135,80)(170,70)
\ArrowLine(120,-2)(85,20)\DashArrowLine(135,20)(85,20){3}\ArrowLine(170,30)(135,20)\ArrowLine(135,20)(170,10)
\Text(-1,39)[c]{$q$}\Text(-1,14)[c]{$\bar q$}\Text(23,33)[c]{\small $ Z'$}
\Text(36,38)[c]{\small $N$}\Text(36,13)[c]{\small $\overline N$}
\Text(71,58)[c]{\small $e^- \, (\nu_e)$}\Text(98,49)[c]{\small $\bar s \; (\bar c\,)$}\Text(94,36)[c]{\small $c$}
\Text(72,-3)[c]{\small $e^+ \, [\bar\nu_e]$}\Text(98,4)[c]{\small $s \; [c\,]$}\Text(94,16)[c]{\small $ \bar c$}
\Text(58,37)[c]{\small $S^+ (S^{0})$} \Text(60,16)[c]{\small $S^- [S^{0\dagger}]$}
\end{picture}

}
\end{center}
\vspace*{-3cm}
\caption{Resonant production of vectorlike neutral leptons, followed by 3-body decays within the $Z'N$ Model (left diagram), or by decays through a scalar doublet $(S^+,S^0)$  within the 
$Z'NS$ Model (right diagram). In both models the final states are $e^+e^-\!+4j$, $e+4j+ \slash \!\!\! \!  E_T$, or $4j+ \slash \!\!\! \! E_T$.}
\label{fig:ZN}
\end{figure}

The process relevant for the $ee$-plus-jets final state is $p p \to Z^\prime \to N \bar{N}$ followed by
$N \to e^- jj$ and $\bar{N} \to e^+ jj$, so that there is a  $e^-e^+ \! + 4j$  resonance at $M_{Z'}$, as shown in the left diagram of Fig.~\ref{fig:ZN}.
The $e\nu$-plus-jets final state arises from the same resonant $N \bar{N}$ production followed by $N \to \nu jj$ 
and  $\bar{N} \to e^+ j j $, or $N \to e^- jj$ and  $\bar{N} \to \bar{\nu} j j $. 
Assuming that the branching fractions for decays of $N$ involving $\mu$, $\tau$, $b$ or $t$ are negligible, the branching fractions for 
$N\to e j j$ and $N\to \nu j j$ are equal to 50\%.
Thus, the total cross section for the $p p \to Z' \to e^+e^- +4j $ and $p p \to Z' \to e \nu  +4j $ processes, for a narrow $Z'$, are
\be
\sigma(ee+4j) = \frac{1}{2} \sigma(e\nu+4j) = \frac{1}{4} \sigma ( pp \to Z' X) B(Z' \to N\bar{N})  ~~.
\label{cross-sections}
\ee
For $M_{Z'} = 2.3$ TeV and $m_N = 700$ GeV we find $\sigma(ee+4j) = g_z^2 \times 1.8 $ fb  at the 8 TeV  LHC.

\subsection{$Z'NS$ Model}

In the second model, referred to as $Z'NS$, instead of the dimension-7 operator (\ref{dim-6}) there is  an $SU(2)_W$-doublet scalar, $S = (S^+, S^0)$, which is neutral under $U(1)_B$
and has a Yukawa-like dimension-5 interaction
\be
y_{S} \frac{\phi_B}{M_{\cal S'}} 
 \bar{\cal L}_L^1 N_R  i\sigma_2  S^* + {\rm H.c.}  ~~,
 \label{eq:NeS}
\ee
where $y_S$ is a dimensionless parameter, and $M_{\cal S'}$ is the mass of a very heavy particle that has been integrated out,
for example an $\cal S'$ scalar of $U(1)_B$ charge $+1$.
The above dimension-5 operator, with $\phi_B$ replaced by its VEV, induces the $N\to S^0 \nu$ and $N\to S^+ e^-$ decays. For simplicity, we take 
the masses of $S^0$ and $S^+$ to be equal, {\it i.e.}, they get contributions only from $SU(2)_W$ invariant terms.
Furthermore, we include a large positive squared mass for $S$ and forbid any term in the scalar potential that is linear in $S$,
so that $S$ has no tree-level VEV. 

We assume that $S^0$ and $S^+$ decay predominantly into a quark-antiquark pair of the second or first generation, via 
dimension-5 operators of the type 
\be
\frac{ \varphi }{m_\psi} \bar{\cal Q}_L^2 c_R  i\sigma_2  S^* + {\rm H.c.}  ~~,
\ee
where $m_\psi$ is the mass, in the multi-TeV range,  of a vectorlike quark $\psi$ that has been integrated out,
and $\varphi$ is a gauge-singlet scalar that gets a VEV much smaller than $m_\psi$. After replacing $\varphi$ by its VEV, this leads to 
small effective Yukawa couplings of the type $S^+ \bar{c}_R s_L$ and $S^0 \bar{c}_R c_L$. Nevertheless, the decays of the $S$ scalars are 
prompt as long as the effective Yukawa couplings are larger than $O(10^{-7})$. 

At 1-loop, the effective Yukawa couplings generate an $S H^\dagger$ term (where $H$ is the SM Higgs doublet)
in the effective Lagrangian, leading to a tiny, inconsequential  VEV for $S^0$.
Couplings to third generation quarks may be suppressed by a $Z_2$ symmetry under which $S$ and $c_R$ are odd; 
the $S^0 \to c \bar{t}$ and $S^+ \to c \bar{b}$ decays are still allowed, but have CKM suppressed amplitudes.
Therefore, $S^0$ and $S^\pm$ decays into a pair of non-$b$ jets  have branching fractions close to 100\%. 

The process  $p p \to Z^\prime \to N \bar{N}$ followed by $N \bar{N} \to S^+ e^- S^- e^+ \to e^+e^-\!+4j$ or
$N \bar{N} \to S^\pm e^\mp S^{0(\dagger)} \nu \to e\nu +4j$ then generates the relevant signals (see right diagram of Fig.~\ref{fig:ZN}).
The branching fraction for each of the $N\to S^0 \nu$ and $N\to S^+ e^-$ decays can each be very close to 50\%; competing decays
such as $N\to e^\pm W^\mp$,  $N\to \nu Z$, $N\to \nu h$ are suppressed by a very small mixing angle squared, while
$N\to \mu S$ and $N\to \tau S$ can be almost forbidden by assigning first-generation lepton number $+1$ to $N$.
Thus, Eq.~(\ref{cross-sections}) is valid for both the $Z'NS$ and $Z'N$ Models.

\subsection{$Z'ES$ Model}

The third model, referred to as $Z'ES$, has the same particle content as the $Z'NS$ Model, except that $N$ is replaced by 
a vectorlike lepton $E$ (of electric charged $-1$). 
The scalar doublet  has an $SU(2)_W$ invariant mass $M_S$, but 
electroweak symmetry breaking splits the masses of $S^0$ and $S^+$ through the $(S^\dagger H)(H^\dagger S)$ term in the Lagrangian. 
For simplicity we take the CP-even and -odd components of $S^0$ to have the same mass.

We impose that $E$ carries the lepton number of the first generation.
The Yukawa-like coupling involving $S$, $E$ and the first generation lepton doublet takes the form
\be
y_{S} \frac{\phi_B}{M_{\cal S'}}  \bar{\cal L}_L^1 E_R  i\sigma_2  S + {\rm H.c.}  ~~,
\ee
similarly to the operator (\ref{eq:NeS}).
For $M_{S^+} > M_{S^0}$, the charged $S$ scalar decays through $S^+ \to S^0 W^+$, or through a virtual $W$ boson into a 3-body final state 
for $M_{S^+} - M_{S^0} < M_W$. To evade constraints on the $\Delta\rho$ isospin-violating parameter it is actually preferable to have 
a small mass splitting within the $S$ doublet, so that we consider only the 3-body $S^+$ decays.
The $S^+$ decay is prompt provided $M_{S^+} - M_{S^0} > O(1)$ GeV. 

We assume that the neutral scalar decays into two gluons through the loop-induced dimension-6 operator 
\be
\frac{c_S
\alpha_s}{m_\chi^2}S H^\dagger   G^{\mu\nu} G_{\mu\nu}  + {\rm H.c.}  ~~,
\ee
where $m_\chi$ is the mass of a heavy particle running in the loop, and 
$c_S$ is a dimensionless coefficient that depends on the loop integral and on the product of $\chi$ couplings to the $S$ and $H$ doublets.
An $S H^\dagger$ term is also generated at one loop, leading as in the $Z'NS$ Model to a tiny $S^0$ VEV.

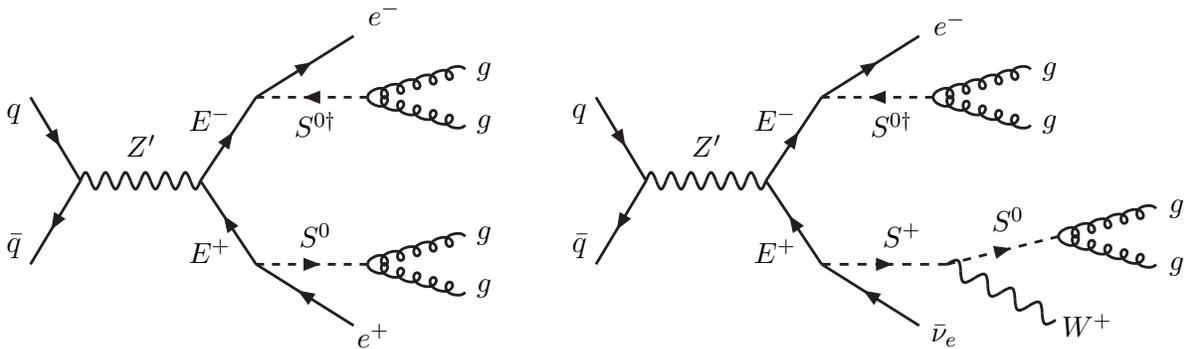
\begin{figure}[b]
\begin{center} 
{
\unitlength=2. pt
\SetScale{1.05}
\SetWidth{1.}      
\normalsize    
{} \allowbreak
\begin{picture}(100,100)(9,-37)
\ArrowLine(4,80)(22,50)
\ArrowLine(22,50)(4,20)
\Photon(22,50)(65,50){3}{6}
\ArrowLine(65,50)(85,80)
\ArrowLine(85,20)(65,50)
\ArrowLine(85,80)(120,102)\DashArrowLine(125,80)(85,80){3}
\Gluon(160,90)(125,80){-2}{5}\Gluon(125,80)(160,70){-2}{5}
\ArrowLine(120,-2)(85,20)\DashArrowLine(85,20)(125,20){3}
\Gluon(125,20)(160,30){2}{5}\Gluon(125,20)(160,10){-2}{5}
\Text(-1,39)[c]{$q$}\Text(-1,14)[c]{$\bar q$}\Text(23,33)[c]{\small $ Z'$}
\Text(36,38)[c]{\small $E^-$}\Text(36,13)[c]{\small $E^+$}
\Text(69,58)[c]{\small $e^- $}\Text(88,47)[c]{\small $g$}\Text(88,37)[c]{\small $g$}
\Text(67,-3)[c]{\small $e^+ $}\Text(88,6)[c]{\small $g$}\Text(88,16)[c]{\small $g$}
\Text(56,37)[c]{\small $S^{0\dagger}$} \Text(56,16)[c]{\small $ S^0$}
\end{picture}
\begin{picture}(100,100)(4,-37)
\ArrowLine(4,80)(22,50)
\ArrowLine(22,50)(4,20)
\Photon(22,50)(65,50){3}{6}
\ArrowLine(65,50)(85,80)
\ArrowLine(85,20)(65,50)
\ArrowLine(85,80)(120,102)\DashArrowLine(125,80)(85,80){3}
\Gluon(125,80)(160,90){2}{5}\Gluon(125,80)(160,70){-2}{5}  
\ArrowLine(120,-2)(85,20)\DashArrowLine(85,20)(130,20){3}\Photon(130,20)(169,0){3}{4}
\DashArrowLine(130,20)(170,30){3}\Gluon(170,30)(205,20){-2}{5}\Gluon(170,30)(205,40){2}{5}
\Text(-1,39)[c]{$q$}\Text(-1,14)[c]{$\bar q$}\Text(23,33)[c]{\small $ Z'$}
\Text(36,38)[c]{\small $E^-$}\Text(36,13)[c]{\small $E^+$}
\Text(69,56)[c]{\small $e^- $}\Text(88,47)[c]{\small $g$}\Text(88,37)[c]{\small $g$}
\Text(68,-3)[c]{\small $\bar\nu_e$}\Text(112,10)[c]{\small $g$}\Text(112,21)[c]{\small $g$}
\Text(58,37)[c]{\small $S^{0\dagger}$} \Text(60,16)[c]{\small $S^+$} \Text(80,19)[c]{\small $S^0$}  \Text(95,-1)[c]{\small $W^+$}
\end{picture}
}
\end{center}
\vspace*{-3cm}
\caption{Resonant production of vectorlike leptons, followed by decays through a scalar $(S^+,S^0)$  within the 
$Z'ES$ Model. The final states are $e^+e^-\!+4j$  (left diagram) or  $e+4j+ W^{(*)} +\slash \!\!\! \!  E_T$  (right diagram).}
\label{fig:ZES}
\end{figure}

The $E^+E^-$ production is followed by the cascade decays shown in Fig.~\ref{fig:ZES}. The $ee+4j$ final state proceeds through 
an intermediate  $S^0$-$S^{0\dagger}$ pair and involves only gluon jets.
The $e\nu+\jets$ final state involves four gluon jets and a (possibly off-shell) $W$ boson. The hadronic decays
of the $W$ boson then lead to a $e\nu+6j$ final state.

The branching fractions of $E^+\to e^+S^{0}$ and $E^\pm\to \nu S^\pm$ are slightly different due to the 
$S^+$-$S^0$ mass difference, while other decay modes of $E^\pm$ can be neglected. 
We obtain
\bear
B(E \to e S^{0}) &=& \frac{(m_E^2 - M_{S^0}^2)^2}{(m_E^2 - M_{S^0}^2)^2 +(m_E^2 - M_{S^+}^2)^2 } 
\\ [3mm]
&\approx & \frac{1}{2} +  \frac{2M_{S}}{m_E^2} \left(M_{S^+} - M_{S^0}\right) + O\left((M_{S^+} - M_{S^0})^2/m_E^2\right) 
\nonumber 
\eear
and $B(E^\pm \to \nu S^\pm) = 1- B(E \to e S^{0})$. 

The branching fractions of $S^0\to gg$ and $S^+ \!\to S^0 W^{+(*)}$ are nearly 100\%. However, the event selection 
discards final states where the $W$ boson (whether on- or off-shell) decays involves an electron or muon,
so the relevant branching fraction is $B(S^+\! \to {\rm hadrons}) \approx 75\%$, 
where we included both $W\to$ jets and $W\to \tau\nu$
with hadronic $\tau$ decays.
Thus, in the $Z'ES$ model, 
\bear
&& \sigma(ee+4j) = \sigma ( pp \to Z' X) B(Z' \to E^+E^-) B(E \to e S^0)^2  ~~,
\nonumber \\ [4mm]
&& \frac{\sigma(e\nu+{\rm hadrons}) } {\sigma(ee+4j) } = 2 \frac{B(E^+ \to \nu S^+) }{B(E \to e S^0)}  B(S^+\! \to {\rm hadrons})  ~~.
\label{cross-sections-ZES}
\eear
For $M_{Z'} = 2.4$ TeV,   $m_E = 800$ GeV,  $M_{S^0} = 400$ GeV and $M_{S^+} = 440$ GeV
we obtain $B(E \to e S^{0})  = 54\%$,
$\sigma(ee+4j) = g_z^2 \times 1.0$ fb  
and $\sigma(e\nu+{\rm hadrons}) = g_z^2 \times 1.3$ fb at the 8 TeV LHC.

\section{Kinematic distributions}
\label{sec:sim}

To study the LHC signatures of the vector-like lepton models described in the previous section, we turn to Monte Carlo simulations. We implement the $Z'N,\, Z'NS$, and $Z'ES$ models into MadGraph 5~\cite{Alwall:2014hca} using the FeynRules~\cite{Alloul:2013bka} package. These models have several new parameters in the form of couplings and masses. We choose the masses of the new particles, as in Table~\ref{tab:charges}, to roughly fit the $eejj$ and $e\nu jj$ excesses.
With these masses set, the overall $pp \to N \bar N, E^+ E^-$ rate can be adjusted by dialing the $Z'$ gauge coupling $g_z$.
The scales and rates of the observed excesses determine the parameters of the $Z'N$ model. In the $Z'NS$, and $Z'ES$, there is additional freedom, in the form of the $S$ mass, and the $S^+-S^0$ mass splitting, to change the morphology of the signal.

\begin{table}[t] \renewcommand{\arraystretch}{1.2}
\begin{center}
\begin{tabular}{|c||c|c|c||c|c|c|} \hline
 new particles & $\;  U(1)_B \;  $ &  $\;   SU(2)_W \;  $  &  $U(1)_Y$  
 & \ $Z'N$ model \ & \ $Z'NS$ model \ & $Z'ES$ model \\
 \hline \hline 
vector-like fermion  $N$     & +1 & 1 & 0 & 700 GeV & 800 GeV & -- \\ [0.3em]  \hline 
vector-like fermion   $E$     & +1 & 1 & $-1$ & -- & --  & 800 GeV \\ [0.3em]  \hline 
scalar  $S = (S^+, S^0)$   & 0 & 2 & +1/2 & -- & 400 GeV & (440, 400) GeV \\ [0.3em] \hline 
gauge boson $Z'$             & 0 & 1 & 0 & 2.3 TeV & 2.4 TeV & 2.4 TeV \\ [0.3em] \hline \hline 
\end{tabular}
\end{center}
\caption{Gauge charges and masses for the new particles included in at least one of the three models described in Section \ref{sec:models}, and involved on-shell in the processes shown in Figs.~\ref{fig:ZN} and \ref{fig:ZES}.}
\label{tab:charges}
\end{table} 

We generate parton-level signal events with MadGraph 5, which are then passed to PYTHIA6.4~\cite{Sjostrand:2006za} for showering, hadronization, and decay. The events are subsequently directed through the DELPHES~\cite{Ovyn:2009tx} package to incorporate detector geometry and response effects. Post-detector level events are then analyzed using the cuts described in Sec.~\ref{sec:selections}.

\begin{figure*}[t]
\centering
\includegraphics[width=0.55\textwidth]{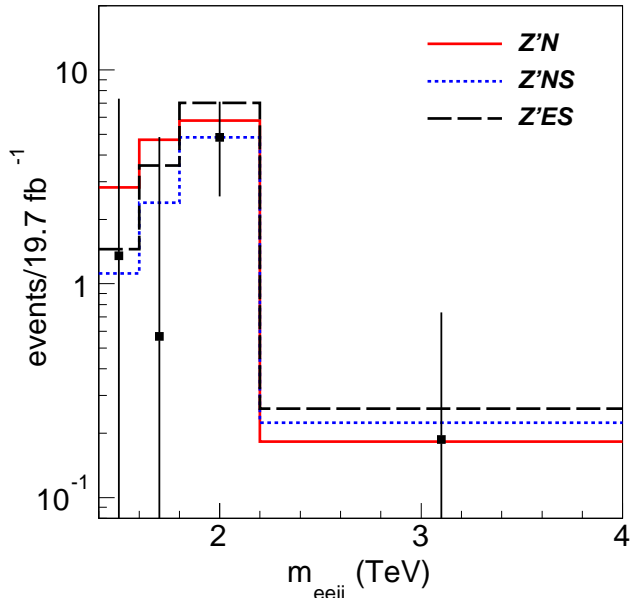}
\caption{The $m_{eejj}$ distribution, formed from the two electrons and two leading jets, for the $Z'N$ (red, solid line), $Z'NS$ (blue, dotted line) and $Z'ES$ (black, dashed line) model. The data points with error bars are taken from~\cite{Khachatryan:2014dka}, after SM background subtraction.
The new particle masses are listed in Table~\ref{tab:charges}. The values of $g_z$ used here are 1.0 for $Z'N$, 1.05 for $Z'NS$, and 1.2 for $Z'ES$.}
\label{fig:meejj}
\end{figure*}

The $eejj$ invariant mass from the models of resonant vector-like lepton production  
with a cross section that roughly fits the CMS data is shown in Fig.~\ref{fig:meejj}. 
The rates used there are given by Eq.~(\ref{cross-sections}) with $g_z =1.0$ for $Z'N$ and $g_z =1.05$ for $Z'NS$, and by Eq.~(\ref{cross-sections-ZES}) with $g_z =1.2$ for $Z'ES$. Note that the distribution has been truncated below $1.5\,\tev$.
 
Comparing this figure with Fig.~2 from Ref.~\cite{Khachatryan:2014dka}, all three models of vector-like leptons describe the data better than a $W'$. The key to the improved fit is the presence of more than two jets in the vector-like lepton signal; the $Z'$ in our signal is actually an $ee+4j$ resonance. As only the leading two jets are included in the $eejj$ distribution, the signal shows an edge, rather than the full Breit-Wigner line shape. As a result, the signal predicts fewer events at high $m_{eejj}$, in agreement with the data. For the $Z'N$ model, the improved $m_{eejj}$ fit is automatic once we have chosen the rate. For the $Z'ES$ and $Z'NS$ models, the agreement depends on the mass of $S$. For $M_S \ll m_E, m_N$, the $S$ scalars emerge boosted from the $N/E$ decay, bringing their decay products (two quarks or gluons) closer together. In events where the $S$ decay products each merge into a single jet, the entire $Z'$ energy will be captured in the electrons and two leading jets; in this case, $m_{eejj}$ has a more characteristic resonance shape, leading to too many events at high $m_{eejj}$. To match the $eejj$ excess in the $1.8-2.2\,\tev$ bin, as well as the agreement between data and the SM in the $2.2-4.0\,\tev$ bin, scalar masses around  $M_S \sim 400\,\gev$ work well. Recall, as discussed in Sec.~III, that $S^0$ lies in an $SU(2)_W$  doublet with the charged state slightly heavier than the neutral one, such that $S^+$ decays to $S^0$ via an off-shell $W^+$.

\begin{figure}[t]
\includegraphics[width=0.49\textwidth]{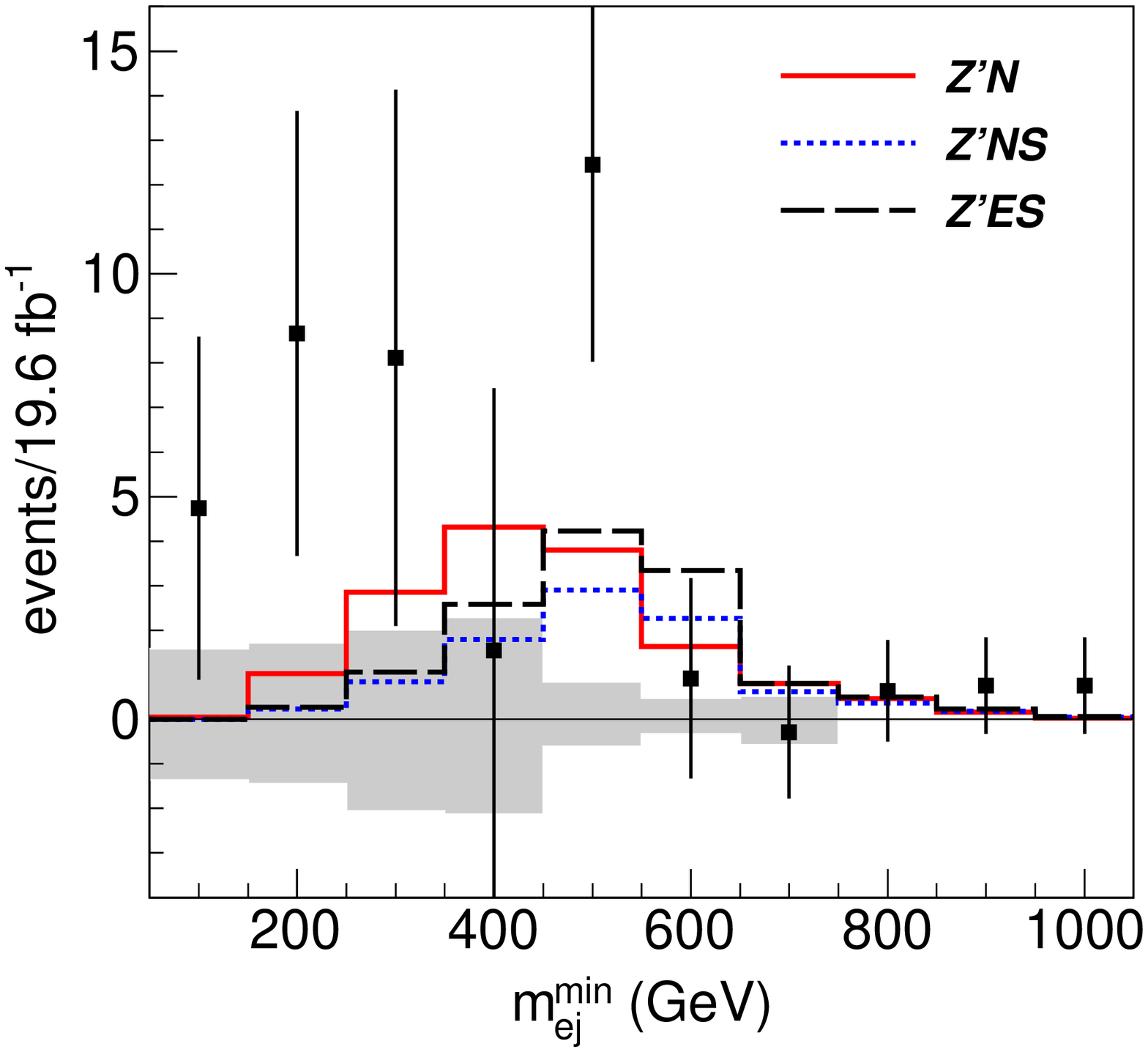}
\includegraphics[width=0.49\textwidth]{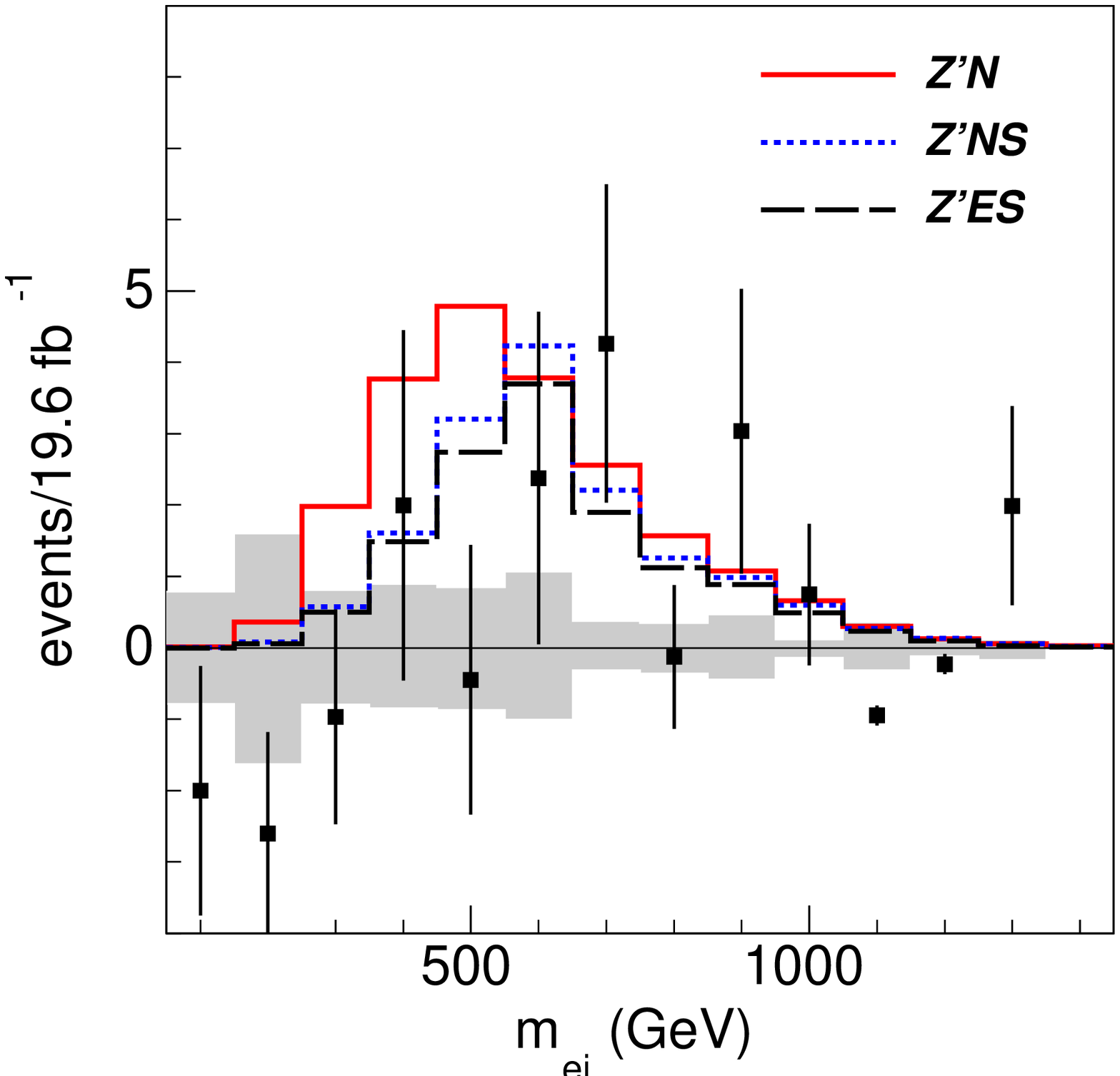}
\caption{The $m^{\text{min}}_{ej}$ (left panel) and $m_{ej}$ (right panel)  distributions for $eejj$ and $e\nu jj$ events, respectively, 
passing the ``leptoquark" selection without  
the $m^{\text{min}}_{ej}$ or $m_{ej}$ cut. The signal colors and parameters are the same as in Fig.~\ref{fig:meejj}. The black data points are taken from~\cite{CMS:2014qpa}, and the shaded histogram is the quoted systematic uncertainty band on the SM background.} 
\label{fig:mej}
\end{figure}

With the signal rate and $S$ mass set to match the $m_{eejj}$ excess of the $W'$ analysis, we now examine how the vector-like lepton signals
appear under the ``leptoquark" event selection.  Applying the leptoquark analysis cuts optimized for a $650\,\gev$ leptoquark, the signal $m^{\text{min}}_{ej}$ and $m_{ej}$ distributions are shown in Fig.~\ref{fig:mej} for the same parameters used in Fig.~\ref{fig:meejj}. 
First, we emphasize that the range of $m^{\text{min}}_{ej}$ and $m_{ej}$ shown in Fig.~\ref{fig:mej} are larger than what CMS uses to determine the significance. As the vector-like lepton models lead to more events in the $200\,\gev < m^{\text{min}}_{ej} < 800\,\gev$ region than the SM, the fit is improved. The improvement of the fit is less clear in the single-electron plus missing energy channel, where the data is lower than the SM at  low $m_{ej}$ and higher than the SM at high $m_{ej}$, transitioning at $m_{ej} \simeq 500\;\gev$. In all cases, the vector-like lepton $m^{\text{min}}_{ej}, m_{ej}$ distributions are wide and do not reveal a sharp peak at $m_N$. This is again due to the presence of extra jets: the $N/E$ decay to an electron (or neutrino) plus a pair of jets and are hence not reconstructed accurately if only one jet is included.

In all of the channels discussed, the statistics are low, so it is possible that the true nature of the excess -- if due to new physics -- requires more data to reveal its shape. As more data will take time to collect, it is worth considering what other distributions to examine in the existing $8\,\tev$ data, both in events with the leptoquark/$W'$ selections and elsewhere, to better gauge the veracity of the excess and its properties. For the vector-like lepton signals discussed here, a first place to look is in the jet multiplicity distribution; for all three models shown here, this distribution peaks at four or more jets. A second distribution is $m_{ejj}$, the invariant mass of either electron when combined with two of the jets in the event, which should reveal a bump at $m_N, m_E$. To more reliably extract an intermediate $N/E$ state, it is useful to require four jets, matching up each electron with two jets such that the two $ejj$ clusters have the smallest mass difference. 
The cleanliness of this approach, though, is reduced by NLO QCD effects, which (especially through initial state radiation) could lead to jets that are more energetic than some of the four jets originating from the $Z'$ decay chain.

Depending on which vector-like model is behind the excess -- should it persist -- there are other signals to be hunted for, both in the existing data and at the future $13\,\tev$ run of the LHC. In all of the models we've discussed, the vector-like leptons are resonantly produced so that a smoking-gun signal for this type of interpretation is a resonant bump in the $ee + 4j$ distribution. Additionally, all models contain a companion signal $\nu\bar{\nu} + \jets$ which should be produced with a rate comparable to that for $ee + 4j$, of the order of $1\,\fb$, and should appear in jets plus $\slashed E_T$ searches. The $Z'NS$ and $Z'ES$ models contain an $S^0$ particle that decays to two jets;  
given that the $S^0\,gg$ coupling is very small, the rates for single and double-$S$ production via gluon fusion are too small to be useful for di-jet or paired-dijet searches. 
A more tractable option is to search for pairs of di-jet resonances {\em within} the $e\nu+\jets$ or $e^+e^- + \jets$ events. 

\setcounter{equation}{0}
\section{\label{sec:Conclusions} Conclusions}

We have presented alternative interpretations of the excess di-electron-plus-jets and electron-plus-jets events recently reported by the CMS Collaboration, in terms of resonantly produced vector-like leptons. While it is possible that the excesses are the result of statistical fluctuations that will shrink with the addition of more data, alternative interpretations are useful; they show what type of new physics would be necessary to describe the data, and provide a list of cross-checks. 
Furthermore, our simple renormalizable models and their possible extensions motivate additional searches. 

The models we explored vary in detail, but all have a set of basic features in common; resonant pair production, through a $Z'$ boson, of a color-singlet
fermion (vector-like lepton) that decay into three or more SM particles. The  decays of the new fermion into a lepton and two or more jets (including $N \to e j j$, $E^+ \to \nu S^+ \to \nu +4j$, etc.) are the key to why these models fit the observed excess better than the CMS signal hypotheses. Regardless of whether the CMS excesses persist, multi-particle resonance searches, including mixtures of leptons and jets like the $ejj$ state studied here, should be added to the toolbox of future LHC analyses.\\ [3mm]

{\bf Acknowledgments}. We would like to thank John Paul Chou, Patrick Fox, Steve Mrenna and Felix Yu for helpful comments and conversations. We are grateful to Pavel Fileviez P\'{e}rez for pointing out 
that the interactions displayed in an earlier version violated $U(1)_B$.
The work of AM was partially supported by the National Science Foundation under Grant No. PHY14-17118. 


\end{document}